\documentclass[12pt]{article}
\usepackage{amsmath,amssymb,amscd}
\usepackage{bm}
\begin{document}
\setcounter{equation}{30}
\subsection*{\it Erratum}

\subsection*{Propagation Effects on the Breakdown of a Linear Amplifier Model: Complex-Mass Schr\"odinger Equation Driven by the Square of a Gaussian Field}

\bigskip
\noindent
{\bf Philippe Mounaix$^1$, Pierre Collet$^1$, Joel L. Lebowitz$^2$}

\bigskip
\noindent
$^1$ Centre de Physique Th\'eorique, UMR 7644 du CNRS, Ecole Polytechnique, 91128 Palaiseau Cedex, France. E-mail: mounaix@cpht.polytechnique.fr; collet@cpht.polytechnique.fr

\noindent
$^2$ Departments of Mathematics and Physics, Rutgers, The State University of new Jersey, Piscataway, NJ 08854-8019, USA. E-mail: lebowitz@math.rutgers.edu

\bigskip
\noindent
Commun. Math. Phys. {\bf 264}, 741-758 (2006)
%
%

\bigskip
\bigskip
\noindent
The proof of the inequality $\lambda_{q}(x,t)\le (q\mu_{x,t} -0^+)^{-1}$ [p 750, below Eq. (29)] is based on the statement that ${\cal E}(x,t;s)$ is an entire function of $s\in {\mathbb C}^M$ [see below Eq. (30)]. But according to Equation\ (9) and Lemma 1, all we know is that ${\cal E}(x,t;s)$ is an entire function of $k(s)\in {\mathbb R}^N$. Nevertheless, the above inequality holds, hence the proposition 1. To prove it we replace (31) with the following lemma.

\bigskip
\noindent {\bf Lemma.}
{\it If $\langle\vert {\cal E}(x,t)\vert^q\rangle <+\infty$, then}
\begin{equation}\label{eq1.1}
\limsup_{\vert\vert s\vert\vert\rightarrow +\infty}{\rm e}^{-\vert\vert s\vert\vert^2}
\vert {\cal E}(x,t;s)\vert^q <+\infty ,
\end{equation}
{\it along almost every direction $\hat{s}$ in ${\mathbb C}^M$.}

\bigskip
\noindent {\it Proof.}
For given $x$ and $t$, write ${\cal E}(x,t;s)=E(\hat{s},\kappa)$ where $\kappa=\vert\vert k(s)\vert\vert =\vert\vert s\vert\vert^2$. Making this change of notation in (29) it is easily seen that if $\langle\vert {\cal E}(x,t)\vert^q\rangle <+\infty$, then
\begin{equation}\label{eq1.2}
\int_0^{+\infty} \vert E(\hat{s},\kappa)\vert^q
{\rm e}^{-\kappa}\kappa^{M-1}d\kappa <+\infty ,
\end{equation}
for almost every direction $\hat{s}$ in ${\mathbb C}^M$. Fix $\hat{s}$ such that\ (\ref{eq1.2}) is fulfilled. By Equation\ (9), Lemma 1, and Lemma 2, $E(\hat{s},z)$ is an entire function of $z\in {\mathbb C}$ of finite exponential type\ \cite{Lev}. Since $q\ge 1$ and $M\ge 1$ are integers, the function $f(z)=E(\hat{s},z)^q{\rm e}^{-z}z^{M-1}$ is also an entire function of $z\in {\mathbb C}$ of finite exponential type, say $\gamma_f$. Let $R$ be a fixed positive number, let $\gamma >\max(0,\gamma_f)$, and define
\begin{equation}\label{eq1.3}
\varphi_{\pm}(z)=
\int_0^R\vert {\rm e}^{\pm i\gamma z}f(z+u)\vert du.
\end{equation}
The functions $\varphi_{\pm}(z)$ are logarithmically subharmonic and bounded by\ (\ref{eq1.2}) on the positive real axis. Furthermore, $\varphi_{+}(z)$ and $\varphi_{-}(z)$ are bounded respectively on the positive and negative imaginary axis. Following then the same argument as in the proof of the Plancherel-P\'olya theorem (see\ \cite{Lev}, p 51), we apply the Phragm\' en-Lindel\"of theorem\ \cite{Lev} to the subharmonic functions $\ln\varphi_{+}(z)$ in the sector $0<{\rm arg}z<\pi /2$ and $\ln\varphi_{-}(z)$ in the sector $-\pi /2<{\rm arg}z<0$. One finds that $\exists A>0$ such that, $\forall z$ with ${\rm Re}(z)>0$,
\begin{equation}\label{eq1.4}
\int_0^R\vert f(z+u)\vert du\le A{\rm e}^{\gamma\vert {\rm Im}(z)\vert}.
\end{equation}
Now, for all $\kappa >R/2$ one has, by the subharmonicity of $\vert f(z)\vert$ and\ (\ref{eq1.4}),
\begin{eqnarray}\label{eq1.5}
\vert f(\kappa)\vert
&\le&\frac{4}{\pi R^2}\int\int_{\vert z\vert <R/2}\vert f(\kappa +z)\vert\, d^2z \nonumber \\
&\le&\frac{4}{\pi R^2}\int_{-R/2}^{+R/2}dy\int_0^R\vert f(\kappa -R/2+iy+u)\vert\, du \\
&\le&\frac{8A}{\gamma\pi R^2}\left({\rm e}^{\gamma R/2}-1\right)<+\infty , \nonumber
\end{eqnarray}
and since $\vert f(\kappa)\vert =\vert E(\hat{s},\kappa)\vert^q{\rm e}^{-\kappa} \kappa^{M-1}$, it follows that, for all $\kappa >R/2$,
\begin{equation}\label{eq1.5b}
\vert E(\hat{s},\kappa )\vert^q{\rm e}^{-\kappa}\le
\frac{8A}{\gamma\pi R^2}\left({\rm e}^{\gamma R/2}-1\right)\frac{1}{\kappa^{M-1}} <+\infty ,
\end{equation}
(hence $\lim_{\kappa\rightarrow +\infty}\vert E(\hat{s},\kappa )\vert^q{\rm e}^{-\kappa}=0$ for $M>1$). Getting back to the original notation yields the new equation\ (\ref{eq1.1}), which completes the proof of the lemma.

\bigskip
We can now proceed with the proof of $\lambda_{q}(x,t)\le (q\mu_{x,t} -0^+)^{-1}$. Since every element of the matrix $\int_0^t\gamma(x(\tau),\tau)\, d\tau$ is a continuous functional of $x(\cdot)\in B(x,t)$ with the uniform norm on $\lbrack 0,t\rbrack$ (see the appendix B), its largest eigenvalue, $\mu_1\lbrack x(\cdot)\rbrack$, is also a continuous functional of $x(\cdot)$. Accordingly, $\forall\varepsilon >0$
$\exists x_{\varepsilon}(\cdot)\in B(x,t)$ such that $\mu_{x,t} -\varepsilon /2\le\mu_1\lbrack x_{\varepsilon}(\cdot)\rbrack\le\mu_{x,t}$. Let $\sigma_{\varepsilon}\in {\mathbb C}^M$ (with $\vert\vert \sigma_{\varepsilon}\vert\vert =1$) be an eigenvector of $\int_0^t\gamma(x_{\varepsilon}(\tau),\tau)\, d\tau$ associated with the eigenvalue $\mu_1\lbrack x_{\varepsilon}(\cdot)\rbrack$. Fix $x_{\varepsilon}(\cdot)$ and $\sigma_{\varepsilon}$. The quadratic form $\hat{s}^\dag\left\lbrack\int_0^t\gamma(x_{\varepsilon}(\tau),\tau)\, d\tau\right\rbrack\hat{s}$ is a continuous function of the direction $\hat{s}$. Thus, $\exists\delta >0$ such that $\forall\hat{s}$ with $\vert\vert\hat{s} -\sigma_{\varepsilon}\vert\vert\le\delta$,
\begin{equation}\label{eq2.1}
\left\vert\hat{s}^\dag
\left\lbrack\int_0^t\gamma(x_{\varepsilon}(\tau),\tau)\, d\tau\right\rbrack
\hat{s} -
\sigma_{\varepsilon}^\dag
\left\lbrack\int_0^t\gamma(x_{\varepsilon}(\tau),\tau)\, d\tau\right\rbrack
\sigma_{\varepsilon}\right\vert\le\frac{\varepsilon}{2},
\end{equation}
and
\begin{eqnarray}\label{eq2.2}
H_{x,t}(\hat{s})&=&\sup_{x(\cdot)\in B(x,t)}\, \hat{s}^\dag
\left\lbrack\int_0^t\gamma(x(\tau),\tau)\, d\tau\right\rbrack
\hat{s} \nonumber \\
&\ge& \hat{s}^\dag
\left\lbrack\int_0^t\gamma(x_{\varepsilon}(\tau),\tau)\, d\tau\right\rbrack
\hat{s} \nonumber \\
&\ge& \sigma_{\varepsilon}^\dag
\left\lbrack\int_0^t\gamma(x_{\varepsilon}(\tau),\tau)\, d\tau\right\rbrack
\sigma_{\varepsilon}-\varepsilon /2\ge \mu_{x,t} -\varepsilon .
\end{eqnarray}
From the latter inequality and Lemma 2 it follows that $\forall\varepsilon >0$, $\exists\delta >0$ such that $\forall\hat{s}$ with $\vert\vert\hat{s} -\sigma_{\varepsilon}\vert\vert\le\delta$,
$$
\limsup_{\vert\vert s\vert\vert\rightarrow +\infty}\frac{\ln\left\vert {\cal E}(x,t;s)\right\vert^q}
{\vert\vert s\vert\vert^2}\ge q\lambda (\mu_{x,t} -\varepsilon) ,
$$
and for every $\lambda >(q\mu_{x,t} -q\varepsilon)^{-1}$,
\begin{equation}\label{eq2.3}
\limsup_{\vert\vert s\vert\vert\rightarrow +\infty}{\rm e}^{-\vert\vert s\vert\vert^2}
\vert {\cal E}(x,t;s)\vert^q =+\infty .
\end{equation}
Since $\delta >0$,  the set of all the directions $\hat{s}$ fulfilling\ (\ref{eq2.3}) is of strictly positive measure and, according to the lemma above, $\langle\vert {\cal E}(x,t)\vert^q\rangle =+\infty$. Therefore, $\lambda_{q}(x,t)\le (q\mu_{x,t} -q\varepsilon)^{-1}$ and taking $\varepsilon$ arbitrarily small one obtains $\lambda_{q}(x,t)\le (q\mu_{x,t} -0^+)^{-1}$. (We use the notation $q\mu_{x,t} -0^+$ to emphasize the fact that $\langle\vert {\cal E}(x,t)\vert^q\rangle$ may be finite at $\lambda =1/q\mu_{x,t}$. Our approach does not yield the behavior of $\langle\vert {\cal E}(x,t)\vert^q\rangle$ at $\lambda =1/q\mu_{x,t}$ sharp.)
%
%

%
%
\end{document}